\documentclass[12pt]{article}
\usepackage{graphicx}
\usepackage{multicol} 
\usepackage{color}

\usepackage{amsmath,amssymb,mathrsfs}

\definecolor{rosso}{cmyk}{0,1,1,0.4}
\definecolor{rossos}{cmyk}{0,1,1,0.55}
\definecolor{rossoc}{cmyk}{0,0.5,1,0.2}
\definecolor{blu}{cmyk}{1,1,0,0.3}
\definecolor{blus}{cmyk}{1,1,0,0.6}
\definecolor{blucc}{cmyk}{1,0.4,0.2,0}
\definecolor{viola}{cmyk}{0,1,0,0.6}
\definecolor{viola2}{cmyk}{0,1,0.2,0.6}
\definecolor{verde}{cmyk}{0.92,0,0.59,0.25}
\definecolor{verdec}{cmyk}{0.92,0,0.59,0.15}
\definecolor{verdes}{cmyk}{0.92,0,0.59,0.4}
\font\tenrsfs=rsfs10 at 12pt
\font\sevenrsfs=rsfs7
\font\fiversfs=rsfs5
\newfam\rsfsfam

\textfont\rsfsfam=\tenrsfs
\scriptfont\rsfsfam=\sevenrsfs
\scriptscriptfont\rsfsfam=\fiversfs
\def\mathscr#1{{\fam\rsfsfam\relax#1}}
\def\Lag{\mathscr{L}}

\newcommand{\SU}{{\rm SU}}
\newcommand{\SO}{{\rm SO}}

\newcommand{\fig}[1]{~\ref{fig:#1}}
\newcommand{\eq}[1]{~{\rm (\ref{eq:#1})}}

\newcommand{\GeV}{\,{\rm GeV}}
\newcommand{\TeV}{\,{\rm TeV}}
\def\circa#1{\,\raise.3ex\hbox{$#1$\kern-.75em\lower1ex\hbox{$\sim$}}\,}

\newcommand{\PRL}{Phys. Rev. Lett.}

\newcommand{\beq}{\begin{equation}}
\newcommand{\eeq}{\end{equation}}
\newcommand{\bea}{\begin{eqnarray}}
\newcommand{\eea}{\end{eqnarray}}
\newcommand{\diag}{\hbox{diag}\,}

\def\circa#1{\,\raise.3ex\hbox{$#1$\kern-.75em\lower1ex\hbox{$\sim$}}\,}
\makeatletter

\def\art{\@ifnextchar[{\eart}{\oart}}
\def\eart[#1]#2#3#4#5#6{{\rm #2}, {\em #3 \rm #4} {\rm (#6) #5} ({#1})}
\def\hepart[#1]#2{{\rm #2, #1}}
\newcommand{\oart}[5]{{\rm #1}, {\em #2 \rm #3} {\rm (#5) #4}}

\newcounter{alphaequation}[equation]
\def\thealphaequation{\theequation\hbox to
0.6em{\hfil\alph{alphaequation}\hfil}}
\def\eqnsystem#1{
\def\@eqnnum{{\rm (\thealphaequation)}}
\def\@@eqncr{\let\@tempa\relax \ifcase\@eqcnt \def\@tempa{& & &} \or
  \def\@tempa{& &}\or \def\@tempa{&}\fi\@tempa
  \if@eqnsw\@eqnnum\refstepcounter{alphaequation}\fi
\global\@eqnswtrue\global\@eqcnt=0\cr}
\refstepcounter{equation} \let\@currentlabel\theequation \def\@tempb{#1}
\ifx\@tempb\empty\else\label{#1}\fi
\refstepcounter{alphaequation}
\let\@currentlabel\thealphaequation
\global\@eqnswtrue\global\@eqcnt=0 \tabskip\@centering\let\\=\@eqncr
$$\halign to \displaywidth\bgroup \@eqnsel\hskip\@centering
$\displaystyle\tabskip\z@{##}$&\global\@eqcnt\@ne
\hskip2\arraycolsep\hfil${##}$\hfil& \global\@eqcnt\tw@\hskip2\arraycolsep
$\displaystyle\tabskip\z@{##}$\hfil
\tabskip\@centering&\llap{##}\tabskip\z@\cr}
\def\endeqnsystem{\@@eqncr\egroup$$\global\@ignoretrue} \makeatother

\newcommand{\sW}{s_{\rm W}}
\newcommand{\cW}{c_{\rm W}}

\begin{document}

\thispagestyle{empty}

%hep-ph/0405040\hfill 
\begin{flushright}
{KA-TP-03-2005\\
IFUP--TH/2005-03\\
hep-ph/0502096\\

}
\end{flushright}
\vspace{1cm}

\begin{center}
{\LARGE \bf \color{rossos} Little-Higgs corrections\\[4mm]
to precision data after LEP2}\\[1cm]

{
{\large\bf Guido Marandella$^a$,  Christian Schappacher}$^b$,
{\large\bf Alessandro Strumia}$^{c}$
}  
\\[7mm]
{\it $^a$ Theoretical Division T-8, Los Alamos National Laboratory, Los Alamos, NM 87545, USA}\\[3mm]
 {\it $^b$ Institut f{\"u}r Theoretische Physik, Universit{\"a}t Karlsruhe, Germany}\\[3mm]
{\it $^c$ Dipartimento di Fisica dell'Universit{\`a} di Pisa
and INFN, 
Italia}\\[1cm]
\vspace{1cm}
{\large\bf\color{blus} Abstract}

\end{center}
\begin{quote}
{\large\noindent\color{blus}
We reconsider little-Higgs corrections to precision data.
In five models with global symmetries SU(5), SU(6), SO(9)
corrections are (although not explicitly) of `universal' type.
We get simple expressions for 
the $\hat{S},\hat{T},W,Y$ parameters, which  summarize all effects.
In all models $W,Y\ge 0$ and in almost all models $\hat{S}>(W+Y)/2$.
Results differ from previous analyses,
which are sometimes incomplete, sometimes incorrect,
and because we add LEP2 $e\bar e\to f\bar f$ cross sections to the data set.
Depending on the model the constraint on $f$ ranges between $2$ and $20\TeV$.\\[2mm] \indent
We next study the `simplest' little-Higgs model 
(and propose a new related model)
which is not `universal'
and affects precision data due to the presence of an extra $Z'$ vector.
By restricting the data-set to the most accurate leptonic data we show
how corrections to precision data generated by a generic $Z'$ 
can be encoded in four effective  $\hat{S},\hat{T},W,Y$ parameters,
giving their expressions.}
\end{quote}

\newpage

%\tableofcontents

%\setcounter{page}{1}
\setcounter{footnote}{0}

\section{Introduction}
While supersymmetric extensions
of the Standard Model (SM) technically solve the hierarchy problem, 
the ever-rising lower bounds on sparticle masses
recently stimulated searches for alternative methods of  electroweak breaking.
% C.A. Scrucca et al., hep-ph/0304220

Models where the Higgs (and possibly other SM particles) become extended objects
seem generically disfavored by precision data~\cite{LEPparadox}, 
which do not show hints for the expected form factors.
Using the QFT language, one expects the presence of
extra dimension 6 operators added to the SM Lagrangian.
Even if new physics is confined to the Higgs sector, precision data are affected
by operators like $|H^\dagger D_\mu H|^2$.
Physically, this happens because experiments tested the $W,Z$ bosons, which contain
Higgs degrees of freedom in their longitudinal components.
Models of this type include technicolor~\cite{tech}, where the Higgs becomes a bound state,
and extra-dimensional models that allow TeV-scale quantum gravity~\cite{add}, 
where the Higgs supposedly becomes some stringy-like object.

\medskip

Attempts of improving the situation employ the fact that
the Higgs mass can be partially protected from 
quadratically divergent one loop corrections
assuming that the Higgs is a pseudo-Goldstone
boson of some global symmetry spontaneously broken at a scale $f$.
In order to address the hierarchy problem in this way $f$
must be around the Fermi scale: 
this has been achieved only recently and only partially by little-Higgs models~\cite{Arkani-Hamed:2001nc,Arkani-Hamed:2002qx,Arkani-Hamed:2002qy,Gregoire:2002ra,Low:2002ws,Chang:2003un,Skiba:2003yf,Chang:2003zn,Perelstein2,Tparity,Schmaltz:2004de}.
The difficulty is that the Higgs does not look like a pseudo-Goldstone boson
(the Higgs has sizable interactions with itself and with the top quark), so that
one has to invent appropriate `epycicles'.\footnote{Little-Higgs models can be
compared to models  with GUT-scale $f$, previously proposed as
solutions to the doublet/triplet splitting problem of supersymmetric unified theories~\cite{SU6}.
In these models Higgs self-interactions and  the top Yukawa coupling arise naturally.
Supersymmetric models need two Higgs doublets: the Goldstone mechanism
forces a flat direction $m_{H_1}^2 + m_{H_2}^2 - 2\mu^2 =0$
without forbidding the usual $D$-term Higgs self-interactions.
RGE running induced by the top Yukawa lifts the flat direction towards
larger $\tan\beta$.
The mechanism employed in little-Higgs models to get the
top Yukawa coupling by adding extra real fermions
is naturally operative in SU(6) unified models because its 20 representation is pseudo-real
(i.e.\ its mass term is forbidden by SU(6) gauge invariance)
and contains one up-type quark~\cite{SU620}.}

%In the past, relatively simple models with GUT-scale breaking allowed
%a  solution to the doublet/triplet splitting problem of supersymmetric unified theories~\cite{SU6}.
%However, in order to address the hierarchy problem in this way $f$
%must be lowered down to the Fermi scale: 

Little Higgs models are mostly characterized by
the choice of gauge and global groups.
The main free parameters are the gauge coupling(s) of the full gauge group,
and the scale $f$ at which the full group is spontaneously broken to the SM group.
No specific model seems better than the other ones.
Tree level exchange of new heavy vector bosons
gives rise to corrections to precision observables.
Such corrections also depend on how fermions are introduced.
We here stick to the simplest choice made in the original literature,
although introducing more `epycicles' gives interesting alternatives~\cite{Tparity}.\footnote{Models with $T$-parity eliminate tree level effects, thereby
allowing $f\sim v$ (provided that UV-divergent loop effects are small).
However a small $v$ is naturally accommodated only under the assumption
that $f$ is also small.  But $f$, like $v$ in the SM, is controlled by mass terms
of Higgs scalars plagued by quadratically divergent one-loop quantum corrections.
Therefore the problem that these models would like to solve is only shifted.}

We improve on previous computations  and analyses~\cite{Csaki:2002qg,Hewett:2002px,Han:2003wu,Gregoire:2003kr,casalbuoni,Kilic:2003mq,Kilian:2003xt} in the following ways.
Most little-Higgs models are `universal': all effects can be
encoded in four parameters, $\hat S,\hat T,W,Y$.\footnote{
 4 parameters are needed, because there are 4 `universal'  dimension 6 operators~\cite{Grinstein,Barbieri:2004qk},
 listed in table~\ref{tab:STUVXYW}.
 Therefore previous analyses  that employed the traditional $S,T,U$ parameters are incomplete~\cite{Peskin}.
 Furthermore in no way the $U$ parameter is a linear combination of $\hat{S},\hat{T},W,Y$.
The $U$ parameter  corresponds to an higher-order dimension 8 operator:
we can ignore all these subleading effects since $f\gg v$.
Beyond adding $W,Y$ we often get values of $S$ and $T$ different from those
obtained in  previous analyses.}
As discussed in section~\ref{LH}
this makes results and computations much simpler than in previous analyses
that tried to compute corrections to all observables.
We include LEP2 data on $e\bar{e}\to f\bar{f}$ cross sections~\cite{LEPEWWG,LEP2}, 
 that provide significant constraints on $W,Y$~\cite{Barbieri:2004qk}.
In section~\ref{exp} we critically discuss the robustness of experimental inputs.

In section~\ref{sec:LittlestHiggs} we consider two `littlest' Higgs models~\cite{Arkani-Hamed:2002qy,Perelstein2},
with global symmetry SU(5) and different gauge groups.
In section~\ref{SO9}  we consider one model with global symmetry SO(9)~\cite{Chang:2003zn}.
In section~\ref{sec:SU6} we consider two little-Higgs models, with global symmetry SU(6)~\cite{Low:2002ws,Gregoire:2003kr}
and different gauge groups.
All these models are `universal'.

In section~\ref{sec:schmaltz} we consider the `simplest' little-Higgs model~\cite{Schmaltz:2004de}.
In section~\ref{sec:oldest} we propose and analyze a related little-Higgs model.
Both models are not universal and affect precision data plus LEP2 
only due to the presence of a heavy $Z'$ boson.
In the previous section~\ref{sec:Z'} we show how, by considering only
 the most precise precision data,
 the effects of generic $Z'$ models can be condensed
in a set of effective $\hat{S},\hat{T},W,Y$ parameters
and compute them.

Section~\ref{concl} contains our conclusions.

\begin{table}
$$\hspace{-4mm}
\begin{array}{rclrlcc}
\multicolumn{3}{c}{\hbox{Adimensional form factors}}&
\multicolumn{2}{c}{\hbox{operators}}\\ \hline
(g'/g){\color{blus}\widehat{S}} &=& \Pi'_{W_3 B}(0) & {\cal O}_{WB}~=&(H^\dagger \tau^a H) W^a_{\mu\nu} B_{\mu\nu} \!\!\!\\[1mm]
M_W^2{\color{blus} \widehat{T} }&=& \Pi_{W_3 W_3}(0)-\Pi_{W^+W^-}(0)\!\!\!
& {\cal O}_H~=&|H^\dagger D_\mu H|^2\\[1mm]
2M_W^{-2}{\color{blus} Y} &=&\Pi''_{BB}(0) &{\cal O}_{BB}~=&(\partial_\rho B_{\mu\nu})^2/2\\[1mm]
2 M_W^{-2}{\color{blus} W} &=& \Pi''_{W_3 W_3}(0) & {\cal O}_{WW} ~=&(D_\rho W^a_{\mu\nu})^2/2\\[1mm]
\end{array}$$
  \caption{\label{tab:STUVXYW}\em 
  The first column defines the adimensional form factors.
The second column defines the {\rm SU(2)$_L$}-invariant universal dimension-6 
operators,
  which contribute to the form-factors on the same row.
  We use canonically normalized fields and inverse propagators $\Pi$.}
\end{table}

\section{Experimental data}\label{exp}
Our data-set includes all traditional precision electroweak data.
Some measurements achieved better than per-mille accuracy.
Most data have per-cent accuracy: LEP2 $e\bar{e}\to f\bar{f}$ cross sections, 
atomic parity violation, M\o{}ller scattering,
neutrino/nucleon scattering, etc.
Despite the larger uncertainty LEP2 plays an important r\^ole:
being the only precision data measured above the $Z$-peak,
LEP2 data are  particularly sensitive to high-energy new physics.

Before performing a global fit, we discuss its `robustness' i.e.\
how necessary arbitrary choices affect the final results.
Since the data-set contains several observables, 
on statistical basis one expects
a few  `anomalous' results. 
Indeed the data contain three $\sim3\sigma$ anomalies.
Only the first one involves one 
measurement which has a significant impact in the global fit.
\begin{itemize}
\item[1)]
There is a $3\sigma$ discrepancy between LEP and SLD
measurements of the weak angle in leptonic couplings of the $Z$.
We do not see how it might be due to new physics.
Assuming that the discrepancy is due to a statistical fluctuation, we
include both pieces of data in our global fit.
%Our global fit contains both data, averaged as dictated by statistics.

\item[2)]
NuTeV claims that the low-energy couplings of neutrinos to left-handed quarks
is $3\sigma$ away from the SM central value.
Hadronic uncertainties have not been fully taken into account in the NuTeV results and
certain SM effects, such as a $s/\bar{s}$ momentum asymmetry~\cite{s}, 
can explain the NuTeV anomaly.
Therefore the fit on which our results are based includes all data except NuTeV
(second row of table~\ref{tab:central}).
\footnote{Ref.~\cite{s} claimed that the NuTeV anomaly cannot be due to `heavy universal' new physics,
but allowing only three free parameters and thereby arbitrarily setting
to zero one linear combination of the four $\hat S,\hat T,W,Y$ parameters.
Nevertheless, by adding LEP2 to the data set, one again finds
that all four parameters are so constrained that
`heavy universal' new physics cannot explain NuTeV.}

\item[3)]
The forward/backward asymmetry in $e\bar{e}\to Z\to b\bar{b}$, $A_{FB}^b$,  is
about $3\sigma$ different from its SM prediction.
It could be produced by new physics, provided that new physics affects almost only $A_{FB}^b$.
In fact $A_{FB}^b$ has an uncertainty much larger than 
other observables, where no anomaly is present.
Furthermore, the total $Z\to b \bar{b}$ rate is in agreement with the SM.
\end{itemize}
We have no reason of dropping $A_{FB}^b$.
Just to verify the stability of our results, in the third row
of table~\ref{tab:central} we study what happens if both NuTeV and $A_{FB}^b$ are dropped.
As both pieces of data mildly favor a heavy Higgs,
omitting them the best-fit value of the Higgs mass decreases below the direct limit $m_h>115\GeV$,
unless physics beyond the SM is present.
This argument was used in~\cite{chanowitz} to claim
that new physics is needed.  % and interpreted as a hint for light sneutrinos in~\cite{A...}.
However the discrepancy has never been significant, and with the most recent data (in particular for the top mass),
the best-fit value of the Higgs mass is only $1\sigma$ too low.
%Adding LEP2 makes the anomaly even weaker.

In conclusion,  constraints on new physics seem `robust'.
On the contrary, possible hints for new physics depend on 
arbitrary choices needed to perform an analysis,
like omitting NuTeV and/or $A_{FB}^b$ and/or adding LEP2.
Since none of these hints is statistically significant
we prefer to ignore them.

\begin{table}$$
\begin{array}{r|cccc|c}
\hbox{Type of fit}
& 1000\,\hat{S} & 1000\,\hat{T}  & 1000\,W & 1000\,Y & \chi^2_{\rm SM} - \chi^2_{\rm min}\\ \hline
\hbox{All data} & -0.3 & -0.6 & -0.7 & 0.4 & 1.1^2\cr
\hbox{Excluding NuTeV (our default fit)} & 0& 0.2 & -0.2 & 0 & 0.5^2\cr
\hbox{Excluding NuTeV and $A_{FB}^b$} & -0.9& -0.3 & -0.4 & 0.2  & 1.2^2\cr
%\hbox{Excluding NuTeV and LEP2} &2.5&1.5 & 1.4 & 1.6  & 1.2^2\cr
\end{array}$$
\caption[X]{\em Best-fit values for $m_h = 115\GeV$. \label{tab:central}
The typical uncertainty on $\hat{S},\hat{T},W,Y$ is $\pm 0.5$ with correlations among them.
The last column shows that in no case the best fit is significantly better 
 than the SM fit.}
\end{table}

\section{Computing universal effects}\label{LH}
We compute the leading effects, suppressed by one power of $v^2/f^2$.
The analysis is simplified by recognizing that, despite appearances, most
little-Higgs models  give corrections of `universal' type, that
can be fully encoded in the four parameters $\hat S,\hat T,W,Y$ defined in~\cite{Barbieri:2004qk}. 
Furthermore,  computations are performed with a simplified technique.
We repeat here the general presentation of~\cite{Barbieri:2004qk} specializing
it to the specific case of little-Higgs models.

There are two sets of vector bosons: charged and neutral.
Each set involves a few vector bosons $W_i$, but all their interactions with the SM fermions
can be written as $J W_1$, where $W_1$ is one linear combination of the $W_i$
and $J$ is the standard SM fermionic current.
This is why corrections are of `universal' type.
The SM contains a few currents $J_\pm, J_3, J_Y$: to get the essential point
we consider a single current and only two vectors $W_{1,2}$.

The mass matrix $m$ of vector bosons $W_i$ receives two contributions:
one related to the scale $f$ of breaking of the full group, and
the usual one related to the scale $v$ of EW symmetry breaking.
The model is built such that the SM vector bosons have  ${\cal O}(v)$ masses,
 while all other ones receive ${\cal O}(f)$ masses.
 The relevant Lagrangian has the schematic form
\beq\label{eq:Kij}\Lag \simeq  W_i \frac{\Pi_{ij}^{\rm full}}{2} W_j+ g JW_1,\qquad  \Pi_{ij}^{\rm full} = p^2 \delta_{ij} - m_{ij}^2.\eeq
The standard computation proceeds by integrating out the heavy mass eigenstates,
which are some linear combination of the $W_i$:
$W_{\rm out} = W_2 \cos\theta + W_1 \sin\theta$.
In this way one obtains the effective Lagrangian for the light mass eigenstate
$W_{\rm in} = W_1 \cos\theta - W_2 \sin\theta$.
The angle $\theta$ is usually determined such that $W_{\rm out}$ is the heavy mass eigenstate.
With more than two vector bosons $\theta$ is replaced by an appropriate unitary matrix.

Let us instead proceed by keeping $\theta$ arbitrary.
Keeping terms up to dimension 6, integrating out $W_{\rm out}$ gives rise to an effective Lagrangian of the form
\beq\Lag_{\rm eff} \simeq W_{\rm in} [A(\theta) + B(\theta) p^2 + C(\theta) p^4] W_{\rm in} + W_{\rm in} [D(\theta) + E(\theta)p^2] J + F(\theta) J^2\eeq
i.e.\ universal corrections $A,B,C$, plus corrections to gauge couplings $D,E$, plus four fermion operators $F$.
At this point one can compute any observable.

%The corrections $D$ and $E$, of apparently `non universal' type,
%are closely related to SM current $J$.
%Using the equation of motion one could rewrite them in terms of explicitely universal 

Alternatively one can recognize that computing all observables
 is not necessary, because 
the apparently `non-universal' terms $E$ and $F$ only involve the SM
current $J$, and can therefore be eliminated by using the equation of motion of $W_{\rm in}$:
$J = - 2 W_{\rm in} [A + B p^2 + Cp^4] /[D+E p^2] $.
The result is an explicitly `universal' effective Lagrangian of the form
with primed coefficients:
\beq\Lag_{\rm eff} \simeq W_{\rm in} [A'(\theta) + B'(\theta) p^2 + C'(\theta) p^4] W_{\rm in} + W_{\rm in} D'(\theta)  J \eeq
With appropriate rescalings of $W_{\rm in}$ and of $g$
it can be rewritten in canonical form
\beq\Lag_{\rm eff} \simeq W_{\rm in} [A''(\theta) + p^2 + C''(\theta) p^4] W_{\rm in} + W_{\rm in} g J .\eeq
Provided that all above steps have been performed correctly one should find that $A''$ and $C''$ do not depend
on the arbitrary angle $\theta$.
Indeed $A''$ and $C''$ have a direct physical meaning,
so that their values cannot depend on how one chooses to compute them.
We do not report the explicit verification of this property
because the needed computations are more cumbersome than illuminating.

This property suggests a simpler way of computing $A''$ and $C''$.
Rather than finding and integrating out the heavy mass eigenstates (which corresponds to one possible choice of $\theta$),
one can more conveniently choose $\theta=0$ and integrate out the vector bosons that do not couple to the SM fermions.
In this way there is no need of diagonalizing the mass matrix, and the 
apparently `non-universal'  terms $E$ and $F$ are not generated.
Therefore all what one has to do is
\begin{enumerate}
\item[1)] Given the model, write down the kinetic matrix $\Pi^{\rm full}_{ij}$ of eq.\eq{Kij}.

\item[2)] Integrate out all the combinations of extra fields not coupled to SM fermions, 
obtaining the effective kinetic term $\Pi$ for 
$W_1 = W^+$ (in the charged sector) or for $W_1=\{B,W^3\}$ (in the neutral sector).
It is given by $\Pi = (\Pi_{\rm full}^{-1}|_{\rm SM})^{-1}$, i.e.\ one has to restrict the inverse 
of the full $\Pi$ matrix
to the fields coupled to the SM currents, and invert it again,
obtaining a $2\times2$ matrix in the neutral sector
and a $1\times1$ matrix in the charged sector.

\item[3)] Expand around $p^2\simeq 0$ and extract $\hat{S},\hat{T},W,Y$.

\end{enumerate}
To be concrete, let us apply this procedure to a model that contains 
one extra heavy vector boson with mass $M$, no mixing to the SM vectors,
and that couples to fermions in the same way as the SM hypercharge. 
The result in the $\{B_\mu,W^3_\mu\}$ basis is
\beq\label{eq:example}\Pi^{-1} = \begin{pmatrix}p^2 - M_W^2 t^2 & M_W^2 t \cr M_W^2 t & p^2 - M_W^2\end{pmatrix}^{-1}
+ \begin{pmatrix} (p^2 - M^2)^{-1} & 0\cr 0 & 0\end{pmatrix}.\eeq
This should be intuitively obvious: the first term is the SM contribution;
the effect of a vector boson that couples like the SM  hypercharge $B_\mu$
is taken into account by adding an extra term to the propagator of $B_\mu$.
From eq.\eq{example} one extracts
$$\hat{S} =Y= \frac{\cW^2}{\sW^2} \hat T = \frac{M_W^2}{M^2},\qquad W=0$$
which, inserted into~\cite{Barbieri:2004qk}
\beq\label{sys:eps123}
\delta \varepsilon_1 \simeq  \widehat{T }-W  - Y \frac{\sW^2}{\cW^2}\, , \qquad
\delta \varepsilon_2 \simeq  -W\, ,\qquad
\delta \varepsilon_3 \simeq  \widehat{S} -W-Y
\eeq
gives
$\delta \varepsilon_{1,2,3}=0$.
Indeed an unmixed hypercharge-like vector affects LEP2 and low-energy observables but does
not affect the traditional precision observables $\varepsilon_{1,2,3}$.

\smallskip

Little Higgs models contain various extra vector bosons of this sort,
that give tree-level corrections to precision observables.
We will study these effects.
Two kinds of extra effects might be relevant.
First, little-Higgs models employ a heavy top quark  to cancel 
the quadratic divergence associated to the top Yukawa coupling. 
This heavy new fermion can give one loop corrections to precision observables mainly through $\hat T$. 
Second, some little-Higgs models also contain Higgs $\SU(2)_L$ triplets with vacuum expectation values,
which can give arbitrary negative corrections to $\hat T$.
In these models we  present the constraint on $f$, computed under two different assumptions:
a) the extra corrections to $\hat T$ are negligible;
b) the extra corrections to $\hat T$ have the value that makes the constraint on $f$ as mild as possible.
%However, given the unknown, and potentially large, contributions to $\hat T$ of Higgs triplets, we choose not to include this contributions in the models where such triplets are present. 
This kind of analysis may well be considered as exhaustive.

The $99\%$ C.L.\ constraints on $f$ are computed at fixed values of the other parameters:
by making the usual Gaussian approximation we impose  $\chi^2(f) = \chi^2_{\rm SM} +6.6$,
which is the value appropriate for 1 degree of freedom.\footnote{
Various previous analyses obtained weaker constraints using the $\Delta\chi^2$
value corresponding
to $n$ degrees of freedom, where $n$ is the number of free parameters 
present in the little-Higgs model under examination. 
Various models have $n=3$, so this is a significant difference.

Indeed, when experiments can determine the allowed range of all $n$ free parameters,
their best-fit range is given in Gaussian approximation by the $n$-dimensional region defined 
by $\chi^2 < \chi^2_{\rm min} +\Delta \chi^2(p)$
where $\Delta \chi^2$ is the value corresponding to $n$ degrees of freedom
at any desired confidence level $p$.
$\Delta \chi^2$ increases with $n$ because
the confidence region has the following meaning:
the joint probability that all parameters lie inside it is $p$.

However, in our case none of the parameters is determined, and
there is only a constraint on $f$.
Our statistical technique is appropriate for this situation,
which is far from the idealized Gaussian approximation.
This is particularly clear in the Bayesian approach to statistical
inference, where $\exp(-\chi^2/2)$ is (proportional to) the density probability
in the parameter space.

Summarizing in a more physical langauge, one should not apply weaker statistical tests
to models that have more unknown parameters.
}
As in~\cite{Barbieri:2004qk} we include all precision data expect NuTeV.

\begin{figure}
\begin{center}
$$\includegraphics[width=8cm]{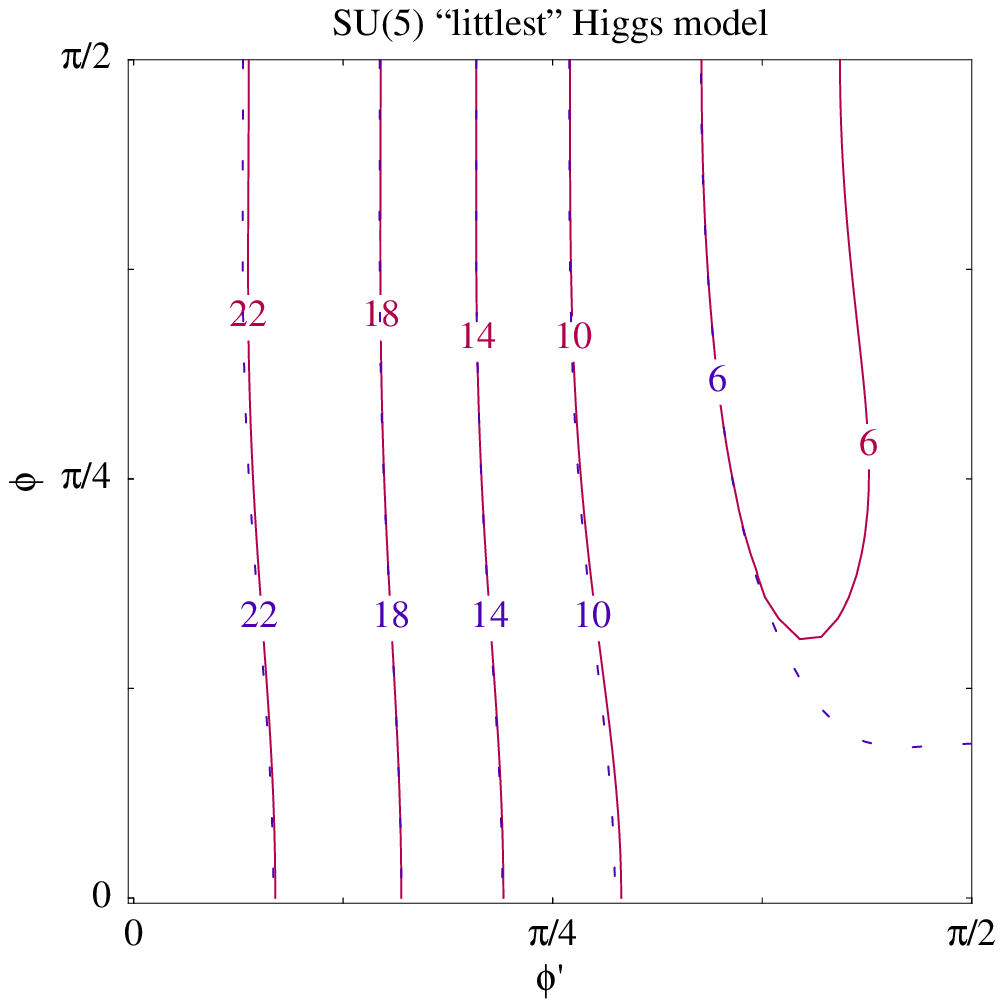}\qquad
\includegraphics[width=8cm]{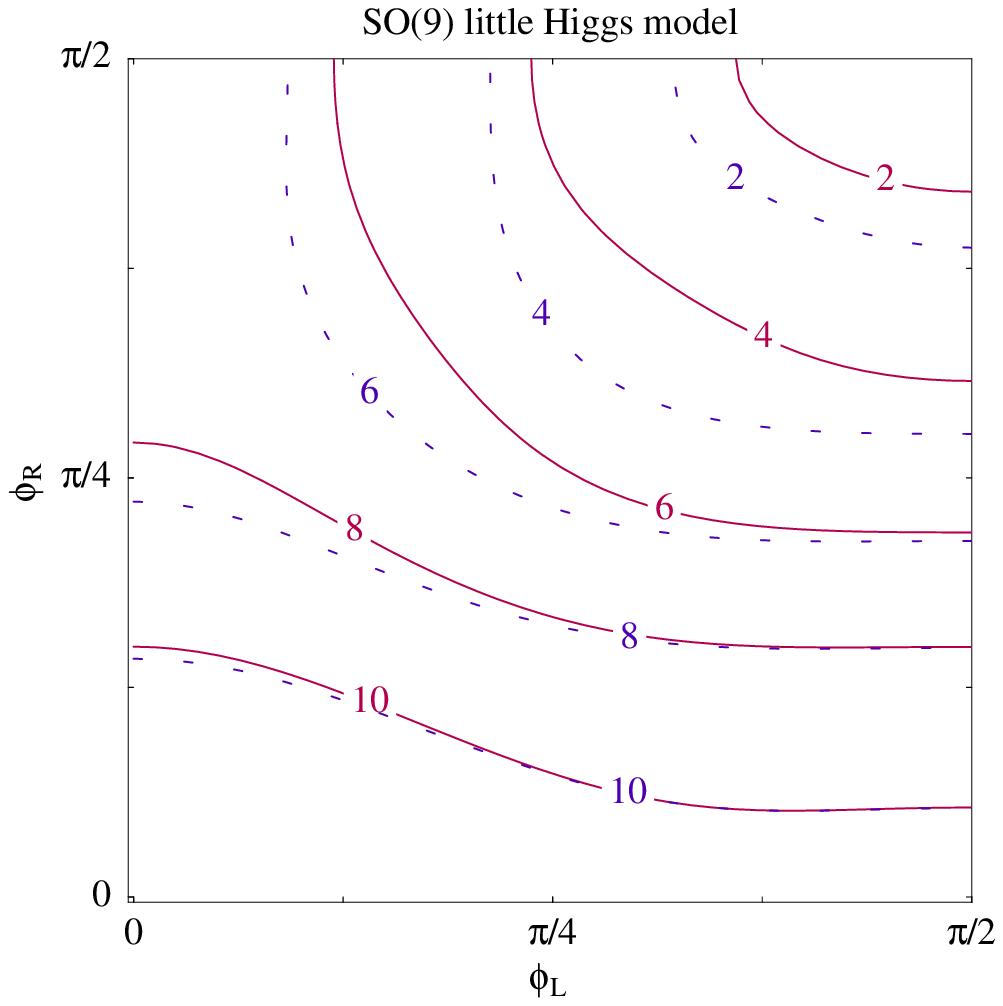}$$
$$\includegraphics[width=8cm]{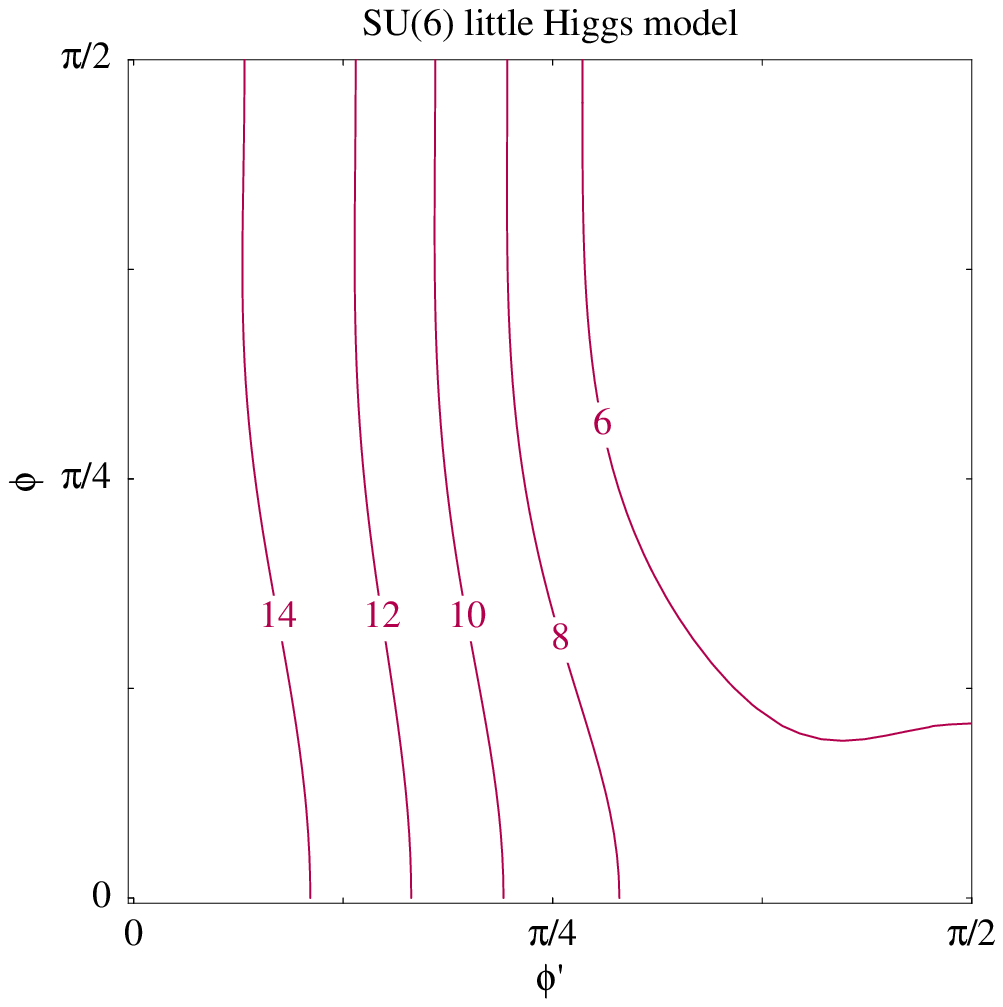}\qquad
\includegraphics[width=8cm]{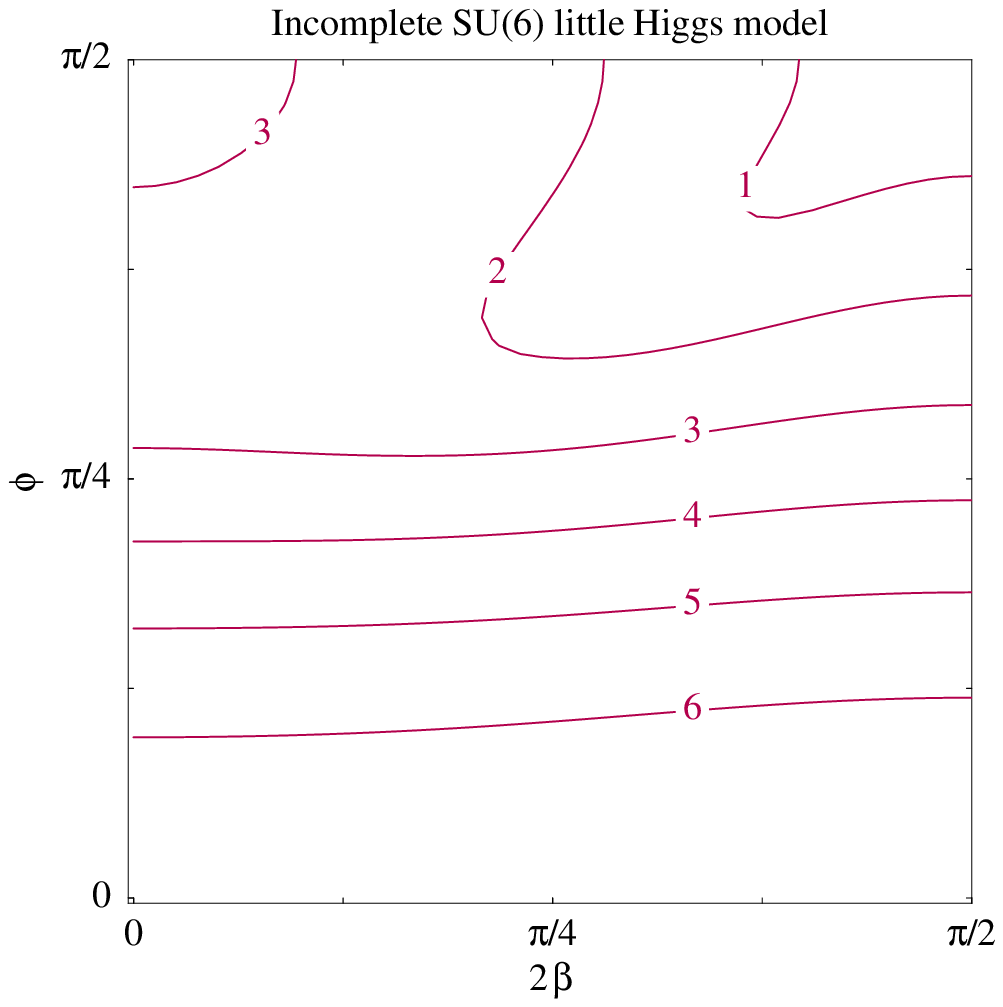}$$
\caption{\em Bound from precision data
on the scale $f$ in {\rm TeV} of little-Higgs models.
The constraint is computed at $99\%$ C.L. for 1 dof, i.e.\
 $\chi^2 = \chi^2_{\rm SM} +6.6$.
As described in the text
in each model the angles $\phi$ parameterize the gauge couplings of the
extra gauge groups, which become strongly coupled at $\phi\to 0$ and/or $\phi\to \pi/2$.
The dotted iso-lines show that the constraint on $f$ gets slightly relaxed
in presence of arbitrary extra corrections to $\hat{T}$.
We assumed a light higgs, $m_h\sim 115\GeV$.
\label{fig:LH}}
\end{center}
\end{figure}

\section{The SU(5)/SO(5) `littlest' Higgs models}\label{sec:LittlestHiggs}
This model has a SU(5) global symmetry broken down to SO(5) at the scale $f$. 
Only the $\SU(2)_1\otimes \SU(2)_2 \otimes {\rm U}(1)_1 \otimes {\rm U}(1)_2$
subgroup of SU(5) is gauged, with gauge couplings
$g_1$, $g_2$, $g'_1$, $g'_2$ respectively.
The SM gauge couplings $g$ and $g'$ are obtained as
$1/g^2 = 1/g_1^2 + 1/g_2^2$ and $1/g^{\prime 2} = 1/g^{\prime 2}_1+1/g^{\prime 2}_2$.
$f$ is normalized such that the heavy vector bosons have masses
\begin{equation}
  \label{eq:heavygb}
  M_{W'}^2 = \left( g_1^2 + g_2^2\right) \frac{f^2}{4}, \hspace{1cm} 
  M_{B'}^2 = \left( g'^2_1 + g'^2_2 \right) \frac{f^2}{20}.
\end{equation}
%When comparing with constraints on $f$ in other models,
%one should take into account that the alternative normalization
%$f\to \sqrt{2} f$ is sometimes adopted.
Matter fermions are charged only under $\SU(2)_1 \otimes {\rm U}(1)_1$.  
See~\cite{Arkani-Hamed:2002qy,Csaki:2002qg} for further details.
The model has three free parameters, which can be chosen to be $f$ and two
angles $\phi$ and $\phi'$ defined by
\begin{equation}
 \sin \phi = g/g_1 \hspace{0.7cm}
  \cos \phi = g/g_2 \hspace{0.7cm}  \sin \phi' = g'/g'_1 \hspace{0.7cm}
  \cos \phi' = g'/g'_2.
\end{equation}
We integrate out the vectors of   $\SU(2)_2 \otimes {\rm U}(1)_2$ under which matter fermions are neutral:
therefore neither 4-fermion operators nor corrections to the SM gauge couplings are generated. 
Only the  propagators of the vector bosons coupled to fermions get modified,
giving rise to
\beq \label{eq:stwylittlest} \begin{array}{ll}\displaystyle
  \hat S = \frac{2 M_W^2}{g^2 f^2} \bigg[\cos^2 \phi + 5\frac{\cW^2}{\sW^2} \; \cos^2
  \phi'\bigg], \qquad&\displaystyle
  W = \frac{4 M_W^2}{g^2 f^2} \cos^4 \phi,\\ \displaystyle
  \hat T = \frac{5 M_W^2}{g^2 f^2}\ ,&
 Y = \displaystyle\frac{20 M_W^2}{g'^2 f^2} \cos^4 \phi'.
\end{array}\eeq
The continuous line of fig.\fig{LH}a shows the $99\%$ C.L. (1 dof) bound on $f$ as function of
$\phi$ and $\phi'$.  
Higgs triplets with a small vev $v_T$ can give an extra negative contribution to $\hat{T}$,
$\hat{T}_{\rm triplets} = -g^2 v_T^2/M_W^2$,
that allows to slightly relax the constraint on $f$ down to the values indicated by the
dotted iso-lines in  fig.\fig{LH}a.
Here and in the following we assume a light higgs, $m_h\sim 115\GeV$.
An acceptable fit with a heavy Higgs $m_h\sim \TeV$ is possible 
in models that give a positive correction to $\hat{T}\sim \hbox{few}\cdot 10^{-3}$.\footnote{Using
the codes in~\cite{FormCalc} we computed how, in the SM at  one-loop, the 
LEP2 $e\bar{e}\to f\bar{f}$ cross sections depend on the Higgs mass $m_h$.
Going from $m_h=115\GeV$ to $m_h=1\TeV$
$\sigma(e\bar{e}\to \mu\bar{\mu}) $ increases by $0.4\%$
and $\sigma(e\bar{e}\to \sum_q q\bar{q}) $ increases by $1.3\%$.
This variation is comparable to experimental uncertainties, so that
LEP2 cross sections do not provide  significant extra informations on $m_h$
beyond $Z$-pole observables.}
In the present model a heavy higgs is allowed for $f\sim 5\TeV$ and appropriate $\phi,\phi'$.

Inserting $\hat{S},\hat{T},W,Y$ into eq.~(\ref{sys:eps123}) 
gives the `littlest'-Higgs contributions to the
$\varepsilon_{1,2,3}$ parameters~\cite{ABC}
%$$\varepsilon_1 = \frac{v^2}{2f^2}(5 - 4\cos^4\phi-20\cos^4\phi'),\qquad
%\varepsilon_2=-2\frac{v^2}{f^2}\cos^4\phi,$$
%$$\varepsilon_3 =- \frac{v^2}{f^2}\bigg[\cos^2\phi\,\cos2\phi
%+5 \frac{\ccW}{\ssW}\cos^2\phi'  \, \cos2\phi'\bigg]$$
that can be compared with~\cite{casalbuoni}: $\varepsilon_1$ and $\varepsilon_3$
do not agree.

\bigskip

The `littlest Higgs' model can be modified by changing the U(1) embedding of the fermions, 
assigning charge $Y R$ under U(1)$_1$ and $Y (1-R)$ under U(1)$_2$, where $Y$ is the SM hypercharge and $0\le R\le 1$~\cite{Csaki:2002qg}.
The corrections to precision data become
% \SU(5)& 32211 & \displaystyle
% \frac{2 M_W^2}{g^2 f^2} \bigg[\cos^2 \phi + 5\frac{\cW^2}{\sW^2}(2R-1) \; (R-\sin^2
%   \phi')\bigg] &\displaystyle
%   \frac{5 M_W^2}{g^2 f^2} (2R-1)^2 +\hat{T}_{\rm triplet}&\displaystyle
%    \frac{4 M_W^2}{g^2 f^2} \cos^4 \phi &
%    \displaystyle\frac{20 M_W^2}{g'^2 f^2} (R-\sin^2 \phi')^2\\[-3mm] \displaystyle
% R=3/5 &  &  & & \\[2mm] \displaystyle
\beq \label{eq:stwylittlest2} \begin{array}{ll}\displaystyle
  \hat S = \frac{2 M_W^2}{g^2 f^2} \bigg[\cos^2 \phi + 5\frac{\cW^2}{\sW^2}(2R-1) \; (R-\sin^2\phi')\bigg],
 \qquad&\displaystyle
  W = \frac{4 M_W^2}{g^2 f^2} \cos^4 \phi,\\ \displaystyle
  \hat T = \frac{5 M_W^2}{g^2 f^2} (1-2R)^2\ ,& \displaystyle
 Y =\frac{20 M_W^2}{g'^2 f^2} (R-\sin^2 \phi')^2.
\end{array}\eeq
The previous case corresponds to $R=1$.
Gauge interactions are not anomalous and are compatible with the needed SM
Yukawa couplings also in the case $R=3/5$~\cite{Csaki:2002qg}. 
For $R=3/5$ the constraint on $f$ becomes slightly weaker than for $R=1$, such that
values of $f$ between 2 and 3 TeV become allowed in some $\phi,\phi'$ range.

\bigskip

The `littlest Higgs' model can also be modified by gauging only $\SU(2)_1\otimes \SU(2)_2 \otimes {\rm U}(1)_Y$~\cite{Perelstein2}. 
In this way $\hat{T}=0$ but one has a quadratic divergence to the Higgs mass associated to the hypercharge coupling $g'$. This model has been already analyzed in terms of $\hat S,\hat T,W,Y$ in~\cite{Barbieri:2004qk} and we report here the results
\beq
  \hat S  = \frac{2 M_W^2}{g^2 f^2} \cos^2 \phi, \qquad
    W  = \frac{4 M_W^2 }{g^2 f^2} \cos^4 \phi,\qquad
  \hat T  =Y  = 0.\eeq
The $99\%$ C.L.\ constraint on $f$ is well approximated by
$f >\max (6.5\cos^2\phi,3.7\cos\phi)\TeV$.
In the limit of small $\cos\phi$ (which corresponds to large $g_2$)
the constraint on $M_{W'}$ approaches a constant.
%$M_{W'}\circa{>}1.2\TeV$

\section{The SO(9)/(SO(5) $\otimes$ SO(4)) model}\label{SO9}
The model,  introduced in~\cite{Chang:2003zn}, is based on the
breaking of a global symmetry SO(9) down to $\SO(5) \otimes \SO(4)$. 
Only the $\SU(2)_L \otimes \SU(2)_R \otimes \SU(2) \otimes {\rm U}(1)$
subgroup of  $\SO(9)$ is gauged, with gauge couplings
$g_L,g_R, g_2,g_1$ respectively. The heavy vector bosons acquire a mass\footnote{Our
$f$ is two times larger than the $f$ defined in~\cite{Chang:2003zn}.
In this way all models are analyzed using the same normalization of $f$
and a clean comparison is possible.
Notice also that we employ the $v=174\GeV$ convention for the SM Higgs vev.}
\begin{equation}
  M_{W'}^2 = \left( g_L^2 + g_2^2\right)\frac{ f^2}{4}, \hspace{1cm} 
  M_{W^{r \pm}}^2 = g_R^2 \frac{ f^2}{4} ,\hspace{1cm} 
  M_{B'}^2= \left( g_R^2 + g_1^2\right) \frac{ f^2}{4}
\end{equation}
and the SM gauge couplings are given by $1/g^2 = 1/g_L^2+1/g_2^2$ and $1/g'^2 = 1/g_R^2+1/g_1^2$.
Matter fermions are charged only under $\SU(2) \otimes {\rm U}(1)$. 
The model has three free parameters, which can be chosen to be $f$ and two
angles $\phi_L$ and $\phi_R$ defined by
\begin{equation}
  \tan \phi_L = g_L/g_2, \hspace{1cm} \tan \phi_R = g_R/g_1 .
\end{equation}
Integrating out the $\SU(2)_L \otimes \SU(2)_R$ vector bosons,
not coupled to fermions,  gives rise to 
\beq\begin{array}{ll}\displaystyle
  \hat S = \frac{2M_W^2}{g^2 f^2} \bigg[ \cos^2 \phi_L +\frac{\cW^2}{\sW^2}\; 
  \cos^2 \phi_R \bigg] ,\qquad &\displaystyle
  W  = \frac{4M_W^2}{g^2 f^2} \cos^4 \phi_L, \\
  \hat T = 0, &\displaystyle
  Y  = \frac{4M_W^2}{g'^2 f^2} \cos^4 \phi_R.
  \end{array}\label{eq:SO9}\eeq
  Higgs triplets can give an extra negative correction to $\hat{T}$.
The continuous line of fig.\fig{LH}b shows the $99\%$ C.L. (1 dof) bound on $f$ as function of
$\phi_L$ and $\phi_R$.  
The dotted line shows the same bound assuming that
an extra correction to $\hat{T}$ makes the constraint on $f$ as mild as possible.
This needs a positive correction to $\hat{T}$, which might arise from 
one-loop exchange of heavy vector-like tops.

Ref.~\cite{Chang:2003zn} studied correction to precision observables by integrating out
the heavy vector boson mass eigenstates, which gives rise to
4-fermion operators together with corrections to gauge boson couplings and to
gauge boson self energies.
Ignoring $W$ and $Y$,~\cite{Chang:2003zn} found,  at leading order in $v^2/f^2$,
$\hat S=0$, $\hat T\neq0$ and a set of non-universal operators.
It should be possible to rewrite the apparently non-universal operators 
as extra corrections to the universal parameters $\hat S,\hat T,W,Y$ such that
the total result agrees with eq.\eq{SO9},
where $\hat S\neq0$  and $\hat T=0$.
Indeed the model was built with a custodial symmetry in order
to avoid corrections to $\hat T$.
Despite this feature, the model is strongly constrained because it affects $\hat{S},W,Y$.

% \subsection{The model \cite{Perelstein:2003wd}}
% \label{sec:Peskin}

% In this model only $\SU(2)^2 \times {\rm U}(1)$ is gauged. The fact that no
% $B'$ is present makes the corrections to $\hat T$ and $Y$ vanishing up to the
% order $v^6/f^6$. However the vev of the Higgs triplet contributes to $\hat
% T$ to the order $v'^2/v^2$:

% \begin{align}
%   \hat S & = \frac{M_W^2 \sin^2 \psi}{g^2 f^2}\\
%   \hat T & = - \frac{g^2 v'^2}{4 M_W^2} \\
%   W & = \frac{2 M_W^2 \sin^4 \psi}{g^2 f^2} \\
%   Y & = \order \left(\frac{v^6}{f^6} \right)
% \end{align}
% where
% \begin{equation}
%   \tan \psi = g_1/g_2
% \end{equation}
% These results are the same obtained in \cite{Barbieri:2004qk} except
% for the contibution of the triplet vev.

 \section{The SU(6)/Sp(6) models} \label{sec:SU6}
 The model, introduced in~\cite{Low:2002ws}, is based on a global symmetry 
 $\SU(6)$ broken down to Sp$(6)$ at the scale $f$. 
The gauge group is $\SU(2)_1 \otimes \SU(2)_2 \otimes {\rm U}(1)_1 \otimes {\rm U}(1)_2$, 
with gauge couplings $g_1,g_2,g'_1,g'_2$,
broken to the diagonal $\SU(2)_L \otimes {\rm U}(1)_Y$ at the scale $f$. 
 The heavy gauge bosons have mass
\begin{equation}
  M_{W'}^2 = \left( g_1^2 + g_2^2\right) \frac{f^2}{4}, \hspace{1cm} 
  M_{B'}^2= \left( g'^2_1 + g'^2_2\right) \frac{f^2}{8}
\end{equation}
and the SM gauge couplings are $1/g^2=1/g_1^2+1/g_2^2$ and $1/g'^2=1/g_1'^2+1/g_2'^2$.
The model contains two Higgs doublets (with vev $v_1$ and $v_2$) and no Higgs triplets.
 For the notations we follow~\cite{Gregoire:2003kr}. 
 We neglect the additional ${\rm U}(1)^2 \to {\rm U}(1)_Y$
 breaking term at a scale $F$ introduced in~\cite{Gregoire:2003kr}. 

If the fermions are charged under $\SU(2)_1 \otimes {\rm U}(1)_1$ one gets:
\beq\begin{array}{ll}\displaystyle
   \hat S = \frac{2 M_W^2}{g^2 f^2} \bigg[\cos^2 \phi +2  \frac{\cW^2}{\sW^2}
   \cos^2\phi' \bigg]\label{eq:S1}  \qquad & \displaystyle
  W  = \frac{4 M_W^2}{g^2 f^2} \cos^4 \phi \\  \displaystyle
   \hat T  = \frac{M_W^2}{2 g^2f^2}  (5+ \cos 4 \beta)&
   \displaystyle
   Y  = \frac{8 M_W^2}{g'^2 f^2} \cos^4 \phi'.
  \end{array}\eeq
 where 
 \begin{align}
   &\tan \beta = v_2/v_1 \hspace{1cm} \cos \phi = g/g_1 \hspace{1cm} \sin \phi = g/g_2 \hspace{1cm} \cos \phi' = g'/g'_1 \hspace{1cm} \sin \phi' = g'/g'_2.
 \end{align}
 The resulting constraint on $f$ is reported in fig.\fig{LH}c, 
assuming $\cos4\beta=0$.
 As clear from the analytical expression, the constraint only mildly depends on $\beta$.
 
 \bigskip
 
 Analogously to the case of `littlest'-Higgs models of section~\ref{sec:LittlestHiggs},
 one can build a related  model by gauging
 only $\SU(2)_1\otimes \SU(2)_2\otimes{\rm U}(1)_Y$.
The model is `incomplete' in the sense that one must
 accept  the quadratically divergent correction to the Higgs mass associated to the small $g'$ coupling. 
In this case 
\beq\begin{array}{ll}\displaystyle
   \hat S = \frac{2 M_W^2}{g^2 f^2} \cos^2 \phi\label{eq:S2}  \qquad & \displaystyle
  W  = \frac{4 M_W^2}{g^2 f^2} \cos^4 \phi  \\  \displaystyle
   \hat T  =\frac{M_W^2}{g^2 f^2} \cos^2 2 \beta &
   \displaystyle
   Y  =0.
  \end{array}\eeq
 The constraint on $f$ in this `incomplete' model is shown in fig.\fig{LH}d,
 and is weaker than in the `complete' model.
The same thing happened for the 'littlest Higgs' model. It is again due to the fact that one gets rid of the (large) contributions of the extra U(1) gauge boson which affects $\hat{S},\hat{T},Y$. 
In this `incomplete' model one still a contribution to $\hat{T}$, 
because the two different Higgs vevs are a source of isospin breaking. 

In these models no Higgs triplet is present, so that we have not 
considered the case of arbitrary $\hat T$.
There is however a one-loop correction (mainly to $\hat T$) of a heavy top-like quark. 
These extra corrections are suppressed by the usual factor $v^2/f^2$ as well as by
a one-loop factor $1/(16 \pi^2)$
and depend on extra parameters, such as the heavy top quark mass and its mixing angle 
with the SM top quark. 
As discussed in~\cite{Gregoire:2003kr} one can find regions of the parameter space where
these extra corrections are negligible.

  \section{Models with generic $Z'$ vector bosons}\label{sec:Z'}
 In the next section we will consider the `simplest' little-Higgs
 models, which gives  non-universal corrections to precision observables,
 due to an heavy extra $Z'$ boson.
 Non-universal effects cannot be fully condensed in $\hat S,\hat T,W,Y$.
 By considering models with a generic non-universal heavy $Z'$ vector boson we here show how
its effects can be approximatively condensed in $\hat S,\hat T,W,Y$.
To this end we consider a reduced set of precision observables
 which includes all most accurately measured observables.
 Therefore the non-universal terms ignored by our approximation
 are not much important.

 \smallskip
 
 A $Z'$ model is characterized by the following parameters:
 the $Z'$ gauge coupling $g_{Z'}$,
 the $Z'$ mass $M_{Z'}$
 and the $Z'$-charges of the Higgs and of the SM fermions:
 $Z'_H$, $Z'_L$, $Z'_E$, etc.
E.g.\  a `universal' $Z'$ would be  a replica of the SM hypercharge
with $Z'_F = Y_F$.%, where the SM  $Y$'s are listed in eq.\eq{Y}.

\medskip
 
The kinetic matrix 
 for the neutral vectors  in the $(B_\mu,W^3_\mu,Z'_\mu$) basis is
$$ \left(\begin{array}{ccc} 
p^2 -M_W^2 t^2 &  M_W^2 t & +g g_{Z'} Z'_H v^2 t \\
  M_W^2 t & p^2 - M_W^2 &  -g g_{Z'} Z'_H v^2 \\
g g_{Z'} Z'_H v^2 t & -g g_{Z'} Z'_H v^2 & p^2 - M_{Z'}^2% - 2 (g_{Z'} Z'_H)^2 v^2  
\end{array} \right) $$
where $t\equiv g'/g$.
We now introduce our approximation.
Rather than integrating out the heavy mass eigenstate,
we integrate out  the combination of gauge bosons that does
not couple to $e_L$ and $e_R$.
This is done by redefining the vector fields
\beq B_\mu \to B_\mu -c_Y Z'_\mu, \qquad
W^3_\mu \to W^3_\mu - c_W Z'_\mu\qquad
c_Y = \frac{g_{Z'} Z'_E}{g' Y_E},\qquad
c_W = \frac{2g_{Z'}}{Y_E g}(Z'_E Y_L - Z'_L Y_E)\eeq
and by eliminating the new $Z'_\mu$ by solving its equation of motion.
One immediately gets:
\beq\label{eq:Z'generic}\begin{array}{ll}\displaystyle
\hat S = \frac{M_W^2}{M_{Z'}^2}(c_W-c_Y/t)(c_W-c_Yt -2 g_{Z'}Z'_H/g),\qquad& \displaystyle
W = \frac{M_W^2}{M_{Z'}^2}c_W^2,\\[5mm]  \displaystyle
\hat{T} = \frac{M_W^2}{M_{Z'}^2}[(c_Y t + 2g_{Z'}Z'_H/g)^2 - c_W^2],& \displaystyle
Y =  \frac{M_W^2}{M_{Z'}^2}c_Y^2.\end{array}\eeq
Notice that $ \hat{T}=0$ whenever $H$ and $L$ have the same
$Z'$ charge, $Z'_L=Z'_H$.
Notice also that $\hat{S}=\hat{T}=0$ whenever $Z'_L+Z'_E+Z'_H=0$, such that
the lepton Yukawa couplings are invariant under the extra ${\rm U}(1)_{Z'}$ symmetry.

\bigskip

The $\hat S,\hat T,W,Y$ defined by this procedure neglect non-universal terms 
that affect fermions $f\neq e_L,e_R$.
We now discuss why this is a good approximation provided that $e,\mu,\tau$ have 
the same $Z'$ charges
and unless the $Z'$ couples 
to quarks much more strongly than to leptons.
Corrections to $M_W,M_Z$, $\mu$-decay and
$Z$-couplings to charged leptons are fully included.
 $Z$ couplings to neutrinos or to quarks are not included, but
 they are measured a few times less accurately than
 $Z$ couplings to charged leptons.\footnote{In the next section we will consider a specific model.
We will check that our approximation is accurate by adding to our simple approximation
also the non universal corrections $\delta g_f$
to on-shell $Z$-couplings to any fermion $f$. The general result is
$\delta  g_f = 2 g_{Z'}^2 M_W^2 Z'_H [Z'_E Y_f-Z'_f Y_E +2 T_{3f}(Z'_E Y_L -Z'_LY_E)]/g^2M_{Z'}^2Y_E$, where $g_f = T_{3f} - \sW^2 Q_f$
is the tree-level SM value.}

Concerning LEP2 we again neglect corrections to quark $Z$ and $\gamma$ couplings.
In view of the higher energy,  LEP2 $e\bar{e}\to f\bar{f}$ cross sections are mainly a probe 
of new four-fermion operators
involving electrons, because electrons are the initial state of LEP2.
All such effects are included in our approximation, which neglects
four-fermion operators involving only quarks and neutrinos, not probed by LEP2.
Low energy observables are less precise than high energy observables.
Our approximation is exact for M\o{}ller scattering,
includes all four-fermion operators that affect atomic parity violation
(but neglects corrections coming from the anomalous $Zf\bar{f}$ couplings, better measured by LEP1),
does not apply to neutrino/nucleon scattering.
We ignore Tevatron constraints on $Z'$ bosons, which are competitive with
precision data only in models where the $Z'$ boson is light ($M_{Z'}\circa{<}500\GeV$)
and has a small gauge coupling ($g_{Z'}\circa{<}0.3$)~\cite{Barbieri:2004qk}.

\section{The `simplest' little-Higgs}\label{sec:schmaltz}
This model is based on an $\SU(3)_c\otimes \SU(3)_L \otimes {\rm U}(1)_X$ gauge group
with gauge couplings $g_3,g,g_X$, broken
down to $\SU(3)_c\otimes\SU(2)_L \otimes {\rm U}(1)_Y$ 
at the scale $f\equiv (f_1^2 + f_2^2)^{1/2}$ by two Higgs triplets
$H_{1,2}$ with vev $\langle H_{1,2}\rangle = (0,0,f_{1,2})$ and 
$X$-charge $-1/3$.
Therefore the unbroken U(1)$_Y$ factor is $Y = X - T_8/\sqrt{3}$
(on triplets $T_8 = \sqrt{3} \,\diag(1/6,1/6,-1/3)$).
The hypercharge gauge coupling is $1/g^{\prime 2} = 1/g_X^2 +1/3g^2$,
and $g$ is the usual $\SU(2)_L$ coupling.

\begin{table}
\begin{center}
\begin{tabular}{|c|ccc|}
\hline
  & $\SU(3)_c$ & $\SU(3)_L$ & ${\rm U}(1)_X$  \\
\hline 
$H_1,H_2$ &1& 3 &$-1/3$\\ 
 $L_{1,2,3}$ & 1 & 3 & $-1/3$\\
$E_{1,2,3}$   & 1 & 1 & 1 \\ 
$U_{1,2,3}$   & $\bar 3$ & 1 & $-2/3$ \\
$D_{1,2,3}$   & $\bar 3$ & 1 & 1/3 \\
$Q_3$& 3 & 3 & 1/3 \\
$Q_{1,2}$ & 3 & $\bar 3$ & 0\\
 \hline
$D'_{1,2}$   & $\bar 3$ & 1 & $1/3$ \\
$U'_3$   & $\bar 3$ & 1 & $-2/3$ \\
 $N'_{1,2,3}$   &1 & 1 & 0 \\
\hline
\end{tabular}\hspace{1cm}
\begin{tabular}{|c|ccc|}
\hline
  & $\SU(3)_c$ & $\SU(3)_L$ & ${\rm U}(1)_X$  \\
\hline
$H$ &1& 3 &$X$\\ 
$\Sigma$ &1& 8 &$0$\\ 
$2\times D_{1,2,3}$   & $\bar 3$ & 1 & $+X$ \\
 $2\times L_{1,2,3}$ & 1 & $\bar 3$ & $-X$\\
 $U_{1,2,3}$   & $\bar 3$ & 1 & $-2X$ \\
$E_{1,2,3}$   & 1 &$ \bar 3$ & $+2X$ \\ 
$Q_{1,2,3}$& 3 & 3 & 0 \\
\hline
\end{tabular}
\caption{\em Charge assignments in the `simplest' little-Higgs model (left) and in the `oldest' little-Higgs model (right).}
\label{tab:model1}
\end{center}
\end{table}

 SM fermions are embedded as follows.
$\SU(2)_L$ singlets become $\SU(3)_L$ singlets, with  $X$-charge $X=Y$.
$\SU(2)_L$ doublets can become either $\SU(3)_L$ triplets 
(with $X = Y +1/6$) or anti-triplets (with $X = Y -1/6$).
The choice is fixed by requiring that the model has no gauge anomalies:
one needs the generation-dependent assignment summarized in table~\ref{tab:model1}.
The extra `primed' fermions are needed to avoid new light fermions.
For more details see~\cite{Schmaltz:2004de,Frampton}.
The model differs from the original model of~\cite{Frampton} by having two
Higgs triplets with the same $X$-charge
which independently break the full gauge group to the SM one:
this implements the  `little-Higgs' mechanism.

%
%Before concluding we remark that the second Higgs triplet,
%needed 
%to implement the little-Higgs mechanism,
%is disfavored by flavour data, which confirmed the
%pattern of flavor violation of the SM with one Higgs doublet.
%Extra unseen flavour effects can be avoided by making ad hoc assumptions~\cite{}.
%We here propose a simpler solution:
%instead of adding a second Higgs triplet one can 
%add a scalar field $\Sigma$ in the adjoint representation of $\SU(3)_L$.
%$\Sigma$ contains a second Higgs doublet which automatically has
%the correct hypercharge, whatever is the $X$-charge of $\Sigma$.
%\xxx{CHECK!}

%}

The model contains
five additional vector bosons: a weak doublet which
neither mixes with the SM gauge bosons nor couples to the SM fermions,
and a weak singlet:
\begin{equation}
  Z'_\mu = s_{Z'} X_\mu + c_{Z'} A^8_\mu, \hspace{1cm} s_{Z'}^2  = t^2/3, \hspace{1cm} t=g'/g.
\end{equation}
Its gauge coupling is given by
$g_{Z'}=g/c_{Z'} =g/\sqrt{1-t^2/3}\approx 0.68$  % 0.60$
and its mass is $M_{Z'}^2=2f^2 g^2/(3-t^2)\approx 0.31f^2$ %0.24 f^2$.
This extra gauge boson both couples to the light fermions and
mixes with the SM neutral vectors.
Light fermions have $Z'$ charges $T_8+\sqrt{3}s_{Z'}^2 Y$, which are not universal.
In particular right-handed leptons $E$ have $Z'$ charge $Z'_E = \sqrt{3} s^2_{Z'}$ 
and $L$ and $H$ have charge
$Z'_L=Z'_H=1/2\sqrt{3}-\sqrt{3}s^2_{Z'}/2$.
%$$Z'_\mu\bigg[ \frac{g^{\prime 2}}{3 g^2 - g^{\prime 2}} (\bar{E}\gamma_\mu E)
%+\frac{g^2 - g^{\prime 2}}{2(3g^2- g^{\prime 2})}( \bar{L} \gamma_\mu L)+\cdots\bigg]$$

We can now apply the approximation for generic $Z'$ models
developed in the previous section.
The `simplest' Higgs model corresponds to
$c_Y = g'/\sqrt{3g^2 - g^{\prime 2}}$
and 
$c_W = -c_Y/t$.
Eq.\eq{Z'generic} 
reduces to \beq\hat S =4W =  \frac{2M_W^2}{f^2 g^2} = \frac{4Y}{t^2},\qquad
\hat T = 0.\eeq
The resulting bound is $f > 4.53\TeV$ at $99\%$ C.L. ($5.2 \TeV$ at  $95\%$ C.L).
According to~\cite{Schmaltz:2004de} atomic parity violation
provides the dominant constraint, $f>1.7\TeV$ at $95\%$ C.L.
Our approximate analysis instead gives a stronger constraint in which
atomic parity violation does not play a significant r\^ole.

We can make our approximate 
analysis more precise by including non-universal corrections to
on-shell $Z$-couplings. They are
$$\delta g_{d_L}=-\delta g_{\nu} = \frac{(g^2-g^{\prime 2}) M_W^2}{2f^2 g^2},\qquad
\delta g_{u_L}=\delta g_{d_R}=0$$
as well as (by construction) $\delta g_{e_L} = \delta g_{e_R}=0$.
Including these effects the bound on $f$ negligibly shifts to $f>4.49\TeV$.

\section{The `oldest' little-Higgs}\label{sec:oldest}
An alternative related
model can be built by embedding the second Higgs doublet
in the adjoint representation $\Sigma$ of $\SU(3)_L$.
The Higgs doublet has the correct hypercharge for any
assignment of the $X$-charge of $\Sigma$.
A light pseudo-Goldstone Higgs is now obtained by suppressing the operator
$H\Sigma\Sigma^* H^*$ instead of $|H_1 H_2^*|^2$.
This modification does not affect 
corrections to precision data.
This related model is a non-unified and non-supersymmetric version
of a pseudo-Goldstone solution to the
doublet/triplet splitting problem studied in~\cite{SU6}.

In the original unified model fermions were embedded in the $15\oplus\bar{6}\oplus\bar{6}$
representation of the unified gauge group of SU(6).
By splitting it into $\SU(3)_c\otimes\SU(3)_L\otimes{\rm U}(1)_X$ fragments of SU(6)
one obtains the alternative embedding described in table~\ref{tab:model1}.
It is interesting to check that it is anomaly free and that it reproduces the known SM fermions;
however these checks are not necessary because these properties are easily verified in the unified version.
The unified embedding differs from the one chosen in the `simplest' little-Higgs model
basically because singlet leptons are extended to $\SU(3)_L$ triplets rather than to $\SU(3)_L$ singlets.
In this way the resulting model can be the low energy limit of an SU(6) unified theory.
(In GUT normalization the $X$ charge in table~\ref{tab:model1} is $X=1/3\cdot \sqrt{3/4}$.
Unfortunately,  unification of gauge couplings fails unless one
considers a `split' supersymmetric version of the model, where only the
super-partners of the higgses are light).

%$\footnote{Unification of couplings constants badly fails,
%roughly fails in the supersymmetric version, 
%and becomes successful in a partly supersymmetric version
%where only the higgs sector is supersymmetric
%(i.e.\ if higgsinos are added to Higgs scalars.  Alternatively one can replicate 6 times
%the non-supersymmetric minimal Higgs content).}).

The  `oldest' little-Higgs model does not have replicated Higgses
and therefore does not have the problems with flavour
typical of multi-Higgs models,
avoided in the `simplest' little-Higgs models 
making ad-hoc assumptions about the Yukawa matrices.
Using the unified matter embedding,
the large top Yukawa coupling can be generated along the lines of~\cite{SU620}: 
by adding to the low energy theory the fragments $(3,\bar{3},X)+\hbox{h.c.}$
of the 20 representation of SU(6).
This is analogous to the heavy vector-like top quark employed by little-Higgs models. 

Corrections to precision data differ only because
right-handed leptons now have a different $Z'$ charge,
$\sqrt{3}(s_{Z'}^2-1)$ rather than $\sqrt{3}s_{Z'}^2$.
Using again our generic expressions valid for a generic $Z'$ heavy boson
we obtain, in the `oldest' little-Higgs model:
\beq  \hat{S} = \hat{T} = W = 0,\qquad
Y = \frac{M_W^2}{2f^2 g^{\prime^2}} (1-t^2)^2.\eeq
Three of the four form factors vanish.
The constraint on $f$ is $f>3.0\TeV$ at 99\% CL.

%
%\section{Models with $T$-parity}
%One can build more complicated models such that there are no computable effects at tree level~\cite{Tparity}.
%This is similar to (and inspired by)  `universal extra dimensions' and has similar problems:
%chiral fermions are obtained at the price of making extra dimensions not universal,
%such that UV-divergent quantum corrections reintroduce the effects
%assumed to be absent at tree level.
%One can hope that unspecified new physics cut-offs UV-divergences at some scale $\Lambda$.
%But if $\Lambda$ is large one-loop effects are not suppressed.
%If $\Lambda$ is small one-loop effects are suppressed and corrections of order
%$\hat{S},\hat{T},W,Y\sim M_W^2/(4\pi f)^2$ would allow to
%lower $f$ down to $f\sim 300\GeV$\footnote{When applied to Higgsless models,
%there is the further problem of understanding why data
%agree with the SM with a light Higgs.},
%but one generically expects extra corrections of order $1/\Lambda^2$.
%E.g.\ in the context of extra dimensions with $1/R\sim 300\GeV$
%there is no reason why higher dimensional operators
%should be suppressed with respect to SM operators,
%since all of them are non-renormalizable.
%In conclusion, we do not understand the interest of this class of attempts.

%
%We view the little-Higgs mechanism as a tool added to 10 TeV technicolor models
%(or to other possible partial solutions to the hierarchy problem)
%in order to make less unnatural EW symmetry breaking at the smaller scale $v$.

   \begin{table}[t]
  $$\begin{array}{cc|cccc}
%  \multicolumn{2}{c|}{\hbox{Symmetry groups}} &
%    \multicolumn{4}{c}{\hbox{`Universal' corrections to precision observables}} \\
\hbox{global} &\hbox{gauge} & \hat S & \hat T &W & Y\\ \hline
\displaystyle
\SU(5)& 32211 & \displaystyle
\frac{2 M_W^2}{g^2 f^2} \bigg[\cos^2 \phi + 5\frac{\cW^2}{\sW^2} \; \cos^2
  \phi'\bigg] &\displaystyle
  \frac{5 M_W^2}{g^2 f^2} +\hat{T}_{\rm triplet}&\displaystyle
   \frac{4 M_W^2}{g^2 f^2} \cos^4 \phi &
   \displaystyle\frac{20 M_W^2}{g'^2 f^2} \cos^4 \phi'\\[2mm] \displaystyle
   \SU(5)& 32211 & \displaystyle
 \frac{2M_W^2}{g^2 f^2} \bigg[ \!\!\cos^2 \phi \!+\!  \frac{\cW^2}{\sW^2} \; 
(\cos^2  \phi'\!-\!\frac{2}{5})\!\bigg] &\displaystyle
   \frac{M_W^2}{5 g^2 f^2} +\hat{T}_{\rm triplet}&\displaystyle
    \frac{4 M_W^2}{g^2 f^2} \cos^4 \phi &
    \displaystyle\frac{20 M_W^2}{g'^2 f^2} (\cos^2  \phi'\!-\!\frac{2}{5})^2 
\\  \displaystyle
\SU(5)& 3221 &
\displaystyle \frac{2 M_W^2}{g^2 f^2} \cos^2 \phi & 0+\hat{T}_{\rm triplet} &
\displaystyle    \frac{4 M_W^2 }{g^2 f^2} \cos^4 \phi&0\\[3mm] \displaystyle
{\rm SO}(9)& 32221    &  
\displaystyle\frac{2M_W^2}{g^2 f^2} \bigg[ \cos^2 \phi_L +\frac{\cW^2}{\sW^2}\; 
  \cos^2 \phi_R \bigg] &0+\hat{T}_{\rm triplet}& 
  \displaystyle\frac{4M_W^2}{g^2 f^2} \cos^4 \phi_L&
  \displaystyle \frac{4M_W^2}{g'^2 f^2} \cos^4 \phi_R\\[3mm] \displaystyle
\SU(6)& 32211 
   & \displaystyle \frac{2 M_W^2}{g^2 f^2} \bigg[\cos^2 \phi +2  \frac{\cW^2}{\sW^2}
   \cos^2\phi' \bigg] & 
   \displaystyle\frac{M_W^2}{2 g^2f^2}  (5+ \cos 4 \beta)&
 \displaystyle   \frac{4 M_W^2}{g^2 f^2} \cos^4 \phi  &
\displaystyle     \frac{8 M_W^2}{g'^2 f^2} \cos^4 \phi'\\[3mm]
\SU(6)& 32211 
     &\displaystyle \frac{2 M_W^2}{g^2 f^2} \cos^2 \phi &
\displaystyle     \frac{M_W^2}{g^2 f^2} \cos^2 2 \beta &
\displaystyle      \frac{4 M_W^2}{g^2 f^2} \cos^4 \phi &0\\
    \SU(3)^2 & 331 
     &\displaystyle\approx \frac{2M_W^2}{f^2 g^2} &\approx 0&
      \displaystyle\approx\frac{M_W^2}{2f^2 g^2} &
      \displaystyle\approx\frac{g^{\prime 2}M_W^2}{2f^2 g^4} \\
    \SU(3)^2 & 331        &\approx 0 &\approx 0&\approx 0 &
      \displaystyle\approx\frac{M_W^2(1-t^2)}{2f^2 g^{\prime 2}} 
\end{array}$$
\begin{center}
\caption{\label{LHS}\em Corrections to precision data in various  little-Higgs models.
The first two columns describe the global and gauge groups.
To fully identify the model one sometimes needs to specify also the fermion content,
which is described in the text.
{\rm 3221} is a shorthand for $\SU(3)\otimes\SU(2)\otimes\SU(2)\otimes{\rm U}(1)$, etc.
$\hat{T}_{\rm triplet}=-g^2v_T^2/M_W^2$ is a possible contribution from Higgs triplets with vev $v_T$.
The last model proposed here can be (but does not need to be) 
the low energy limit of a $\SU(6)$ unified model.
}
\end{center}
\end{table}%

\section{Conclusions}\label{concl}
We studied the corrections to precision data generated in various little-Higgs models 
by recognizing that they are of `heavy universal' type:
all effects can be encoded in four parameters, $\hat S,\hat T,W,Y$.
 Their computation is straightforward, if one
 integrates out vector bosons not coupled to the SM fermions,
 rather than heavy mass eigenstates.
Our results are summarized in table~\ref{LHS},
which simplifies, complements and often corrects previous analyses.
We usually get stronger constraints also because we include LEP2 data, 
that have a significant impact.
The text and fig.\fig{LH} describe the  parameter space allowed by precision tests.
We assumed a light higgs, $m_h\sim 115\GeV$.
Models that give a positive correction to $\hat{T}\sim \hbox{few}\cdot 10^{-3}$
also allow an acceptable fit with a heavy higgs, $m_h\sim \TeV$,
for appropriate values of $f\sim\hbox{few TeV}$ and of the other parameters.

\medskip

The last model of table~\ref{LHS}  is not universal, but precision data
are affected only by the presence of an extra specific $Z'$ vector.
In section~\ref{sec:Z'} we discussed how the effects of a generic extra $Z'$ vector
can be approximatively encoded in a set of $\hat S,\hat T,W,Y$ parameters
by restricting data to processes involving charged leptons,
which presently are the best measured processes.
Applying the general result of eq.\eq{Z'generic} gives the last row of table~\ref{LHS}.

 \medskip
 
 Fig.\fig{LH} shows that the typical constraint is $f> \hbox{few TeV}$:
%the above little-Higgs models give tree level corrections to precision observables
%and therefore
all above little-Higgs models need an uncomfortably high  fine-tuning, 
roughly given by $(f/v)^2\sim 10^3$,
in order to break the EW symmetry at a scale $v\ll f$.\footnote{Fine-tuning has been recently 
studied in~\cite{lastminute} assuming $f=1\TeV$.}
 Fine-tuning decreases
in regions of the parameter space where gauge couplings are large
and other effects become out of control.
Therefore a clean discussion of this issue seems not possible.
As discussed in~\cite{LEPparadox} models where
the  same scale $\Lambda$ suppresses higher-order operators
and cuts-off quadratically divergent corrections to the squared Higgs mass
suffer a  `little hierarchy problem'.
The analyzed little-Higgs models, rather than solving this problem,
provide a specific realization where $\Lambda$ is identified with $f$.

%All proposed solutions to the Higgs mass hierachy problem suffer some fine-tuning. 
%In all cases one can justify it by inventing appropriate antrophic landscape models,
%or one can try to invent less fine-tuned more complicated models.

\bigskip

The constraints on $f$ shown in fig.\fig{LH} can be compared 
with the sensitivity of the future LHC collider.
To conclude we discuss how precision measurements
at a future $e\bar{e}$ collider can test little-Higgs models.
Virtual effects of universal heavy new physics are fully described by the four parameters $\hat{S},\hat{T},W,Y$.
A large set of observables can test the universality hypothesis.
More precise determinations of $\hat{S},\hat{T},W,Y$ would arise from 
LEP1-like measurements around the $Z$-pole.
We remark that more precise determinations of $W,Y$ would also arise from
LEP2-like measurements of $e\bar{e}\to f\bar{f}$ at higher energies.
E.g.\ at energies $E\gg M_Z$ the effect of $W$ is:
$$\frac{\delta \sigma(e\bar{e}\to\mu\bar\mu)}{\sigma(e\bar{e}\to\mu\bar\mu)}\simeq
-\frac{2E^2 W}{M_W^2} \frac{\cW^2}{1+24\sW^4},\qquad
\frac{\delta \sigma(e\bar{e}\to\sum q\bar{q})}{\sigma(e\bar{e}\to \sum q\bar{q})}\simeq
-\frac{2E^2 W}{M_W^2} \frac{\cW^4}{1-2\sW^2+64\sW^4/9}.$$
Unfortunately most little-Higgs models have four free parameters and therefore do not
make univocal predictions for the four observables $\hat{S},\hat{T},W,Y$.
Nevertheless, the explicit expressions in table~\ref{LHS} allow to
derive testable inequalities, such as $\hat{S} \ge (W+Y)/2$
and $\hat{S},\hat{T},W,Y\ge 0$.
Only the second and last model in table~\ref{LHS} violate some of these relations.
The possible vanishing of $\hat{T}$ and/or $Y$ is closely related to the gauge group.

%$\delta\sigma/\sigma$ can be measured with error $1/\sqrt{N}$, where $N$
%is the number of events. 
%Summing eq.s~(\ref{sys:sfermions},\ref{sys:Higgs},\ref{sys:gahi})
%SUSY effects can be approximated as
%$$ W \approx +\frac{\alpha_2 M_W^2}{10\pi}\bigg[\frac{1}{8m_L^2}+
%\frac{3}{8 m_Q^2}+\frac{1}{24 m_A^2}+\frac{1}{3\mu^2}+\frac{2}{3M_2^2}\bigg] .$$

%\paragraph{Acknowledgments}
%We thank G.\ Altarelli, P.\ Gambino for useful discussions.

\frenchspacing\small
%\begin{multicols}{2}

%\end{multicols}
\end{document}